\documentclass[prd,preprint,nofootinbib,showpacs]{revtex4}

\usepackage{epsf,epsfig,subfigure,graphicx,amsmath,amssymb}

\usepackage{amsfonts}
\usepackage{latexsym}
\usepackage{color}

\newcommand{\dis}[1]{\begin{equation}\begin{split}#1\end{split}\end{equation}}
\newcommand{\be}{\begin{equation}}
\newcommand{\ee}{\end{equation}}
\def\bea{\begin{eqnarray}}
\def\eea{\end{eqnarray}}

\newcommand{\eq}[1]{Eq.~(\ref{#1})}

\newcommand{\bfrac}[2]{{\left(\frac{#1}{#2} \right)  }}\newcommand{\VEV}[1]{\langle #1 \rangle}

\newcommand\tev{\,{\rm TeV}}
\newcommand\gev{\,{\rm GeV}}

\newcommand\kev{\,{\rm keV}}
\newcommand\ev{\,{\rm eV}}

\newcommand\treh{T_{\rm R}}     
\newcommand\Tdec{T_{\rm dec}}   

\newcommand\fa{f_{a}}

\newcommand\axino{{\tilde{a}}}
\newcommand\maxino{{m_{\axino}}}

\newcommand\abunda{\Omega_{\axino}h^2}

\def\slash#1{\not\!#1}

\begin{document}

\title{X-ray line signal from decaying  axino warm dark matter }

\author{Ki-Young Choi}
\email{kiyoungchoi@kasi.re.kr}
 \affiliation{Korea Astronomy and Space Science Institute,
  Daejon 305-348,  Republic of Korea}

\author{Osamu Seto}
 \email{seto@physics.umn.edu}
 \affiliation{
 Department of Life Science and Technology,
 Hokkai-Gakuen University,
 Sapporo 062-8605, Japan
}

%

\begin{abstract}
We consider axino warm dark matter in a supersymmetric axion model with R-parity violation.
In this scenario, axino with the mass $m_\axino\simeq 7$ keV can
 decay into photon and neutrino
resulting in the X-ray line signal at $3.5\kev$, which might be the origin
 of unidentified X-ray emissions from galaxy clusters and Andromeda galaxy detected by the XMM-Newton 
X-ray observatory.
\end{abstract}


\preprint{HGU-CAP-030} 

\vspace*{3cm}
\maketitle


\section{Introduction}
\label{introduction}

Various astrophysical and cosmological observations provide convincing evidences for the existence of dark matter (DM). Dark matter distribution spans in wide range of scales from galaxy to clusters of galaxies and the large scale structure of the Universe. 

Recently, anomalous X-ray line emissions have been observed from galaxy clusters and also in the Andromeda galaxy~\cite{Bulbul:2014sua,Boyarsky:2014jta}.
While those might be a result of systematic effects, it would be interesting if the line came from the new source of astrophysical phenomena or from new physics. 
It was suggested that the signal might come from decaying dark matter with the mass and lifetime,
\dis{
m_{\rm DM} &\simeq 7\kev,\\
\tau_{\rm DM} & \simeq 2\times 10^{27}  - 2\times 10^{28} \, \sec, 
\label{xray}
}
assuming that they are the dominant component of dark matter.
Some theoretically interesting particle models have been suggested such as sterile neutrino~\cite{Bulbul:2014sua,Boyarsky:2014jta,Ishida:2014dlp,jian:2014gza}, exciting dark matter~\cite{Finkbeiner:2014sja}, millicharged dark matter~\cite{Aisati:2014nda,Frandsen:2014lfa}, axion like particle~\cite{Higaki:2014zua,Jaeckel:2014qea,Lee:2014xua}, in the effective theory~\cite{Krall:2014dba}.

In this paper, we study the warm dark matter axino in a R-parity violating supersymmetric model.
With bilinear R-parity breakings, neutralinos mix with neutrinos and thus the axino can decay into photon and neutrino. We find that the axino mass with $7$ keV can have the proper lifetime and relic density for the X-ray line emission. In this scenario, as an interesting consequence, the upper bound on the neutrino mass imposes that the Bino mass is lighter than about $10$ GeV.

In Section~\ref{axino} we introduce the model of axino dark matter and in Section~\ref{Rparity} we consider the R-parity violation and decay of axinos.  We summarize in Section~\ref{conclusion}.

\section{axino dark matter}
\label{axino}

The strong CP problem and the hierarchy problem in the Standard Model can be naturally solved in the supersymmeric axion model~\cite{KimRMP10,Nilles84}. If axino, the fermionic superpartner of axion, is the lightest supersymmetric particle (LSP), then it is a good candidate of dark matter~\cite{RTW91,CKR00,CKKR01,Choi:2011yf,Choi:2013lwa}.
The effective operator of the axino can be derived by the supersymmetric transformation of the axion interactions and is given by
\begin{equation}
{\mathcal L}^{\rm eff}_\axino=i\frac{\alpha_s}{16\pi \fa}\overline{\axino}
\gamma_5[\gamma^\mu,\gamma^\nu]\tilde{G}^b
G^b_{\mu\nu}  + i\frac{\alpha_YC_{aYY}}{16\pi \fa}\overline{\axino}
 \gamma_5[\gamma^\mu,\gamma^\nu] \tilde{Y}Y_{\mu\nu}, 
\label{eq:Laxino}
\end{equation}
 where $\fa$ is the Peccei-Quinn breaking scale, and 
 $\alpha_s$ ,$\tilde{G}$,  $G_{\mu\nu}$ and  $\alpha_Y$, $\tilde{Y}$, $Y_{\mu\nu}$ are the gauge couplings, gaugino fields and the field strength
 for $SU(3)_c$ and $U(1)_Y$ gauge groups respectively.
The mass of axino is expected to be of the order of gravitino mass, but it can be much smaller~\cite{Tamv82,Frere83,Masiero84,Moxhay:1984am,ChunKN92,Goto:1991gq}, or much larger~\cite{Chun:1995hc} than the typical supersymmetric particle mass scale,
 depending on the specific models~\cite{Kim:2012bb}.
Here we take the light axino mass of the order of keV as dark matter component.

The primordial axinos are generated from the thermal plasma during reheating after the primordial inflation.
If the reheating temperature is lower than the decoupling temperature~\cite{RTW91}
\begin{equation}
\Tdec= 10^{11}\,\gev \bfrac{\fa}{10^{12}\gev}^2\bfrac{0.1}{\alpha_s}^3,
\label{eq:DecTemp}
\end{equation}
the axinos cannot reach the thermal equilibrium.
Then axinos are generated through scatterings and decay of heavy particles in the thermal plasma, and
 the amount could be abundant enough for axino to be the dominant dark matter component~\cite{CKR00,CKKR01,Choi:2011yf}.
The abundance of thermally produced axinos depends on the reheating temperature for the KSVZ axion model~\cite{KSVZ79}~\footnote{For the DFSZ axion model~\cite{DFSZ81} the axino abundance is almost independent of the reheating temperature in the wide range~\cite{Chun:2011zd,Bae11,Bae:2011iw}.}.
The axino number density to entropy density ratio is estimated as~\cite{CKKR01,Steffen04,Strumia10,Choi:2011yf}
\begin{equation}
Y_\axino = 2.0 \times 10^{-5} g_s^6 \log \bfrac{1.108}{g_s}\bfrac{10^{11}\gev}{\fa}^2\bfrac{\treh}{10^6\gev}.
\end{equation}
With this, the relic density of non-relativistic axino at present is given by
\dis{
\abunda=0.28\bfrac{\maxino}{10\kev}\bfrac{Y_\axino}{10^{-4}}.
}
We can find that $\mathcal O(1-10\kev)$ axino can be a natural candidate for warm dark matter when the reheating temperature is around $10^6 \sim 10^7 \gev$ and Peccei-Quinn scale $\fa=10^{11}\gev$. 
Such thermally produced keV axino is a warm dark matter candidate~\footnote{For non-thermally produced warm axino dark matter, see, e.g., Ref~\cite{Seto:2007ym}.}
 and may solve various problems at the small scale in cold dark matter model~\cite{Klypin:1999uc,Moore:1994yx}.
This range of reheating temperature is free from  the gravitino problem~\cite{Jedamzik:2004er,kkm04}.

\section{R-parity violation and axino decay }
\label{Rparity}

We consider the  bilinear type R-parity violation with the usual $\mu$ -term superpotential in the minimal supersymmetric standard model
\dis{
 W_{\slash{R}_p} = \epsilon_i\mu L_i H_u,\label{Wp}
}
where $L_i$ and $H_u$  are chiral super fields of the lepton doublet and up-type Higgs doublet and $\epsilon_i$ parameterizes the size of the R-parity violation.
By redefining the $L_i$ and $H_d$, we can eliminate the R-parity violating term in~\eq{Wp}, then R-parity violating effect appears only in the scalar potential~\cite{Barbier:2004ez,Hirsch:2004he},
\dis{
V_{\slash{R}_p} = m^2_{L_i H_d} \tilde{L}_iH_d^* + B_i \tilde{L}_i H_u + \rm{h.c.},
}
where the coefficients are $B_i\simeq -B \epsilon_i$ and $m^2_{L_i H_d} \simeq (m^2_{\tilde{L}_i}-m^2_{H_d})\epsilon_i$.
From this scalar potential,  the sneutrinos obtain non-zero vacuum expectation values (VEVs) 
\dis{
\VEV{\tilde{\nu}_i} = - \frac{m^2_{L_i H_d} \cos\beta +B_i \sin\beta}{m^2_{\tilde{\nu}_i}} v,
}
where $\tan\beta\equiv \VEV{H_u}/\VEV{H_d}$ and $v\equiv \sqrt{\VEV{H_u}^2 + \VEV{H_d}^2}/\sqrt2\simeq 174\gev$ and $m_{\tilde{\nu}_i}$ is the sneutrino mass.
Since the non-zero VEVs of sneutrinos induce mixings between leptons and gauginos, the neutrinos mix with neutralinos and can obtain mass at the tree level as~\cite{Hirsch:2000ef,Hirsch:2004he}
\dis{
m_\nu \simeq \frac{\mu^2M_1^2(M_1 g^2 +M_2g^{'2})  /g^{'2} \sum_i  \xi_i^2 }{ (g^2 M_1 +g^{'2} M_2)v^2 \mu \sin\beta\cos\beta - 2M_1M_2 \mu^2} , \label{mnu}
}
 where $\xi_i$ parameterizes the R-parity breaking given by
\dis{
\xi_i =\frac{ g' \VEV{\tilde{\nu}_i} }{\sqrt2 M_1}.
}

The upper bound of the neutrino mass $m_\nu \lesssim 1\ev$ constrains the R-parity breaking parameters.

The baryon asymmetry can be erased in the early Universe by the $B-L$ violating interactions~\cite{Campbell:1990fa,Fischler:1990gn,Dreiner:1992vm}. However if one of the lepton flavors of the R-parity violating couplings is preserved  and the slepton mixing is small enough, the problem can be avoided~\cite{Endo:2009cv}. 

 Due to the R-parity violation, the stability of LSP is not guaranteed anymore~\cite{Hooper:2004qf,Chun:2006ss,Endo:2013si}. 
 For bilinear R-parity breakings, the light axino decays dominantly into photon and neutrino and the decay rate is given by~\cite{Covi:2009pq,Choi:2010jt,Bobrovskyi:2010ps}
 \dis{
 \Gamma_\axino=\sum_i\Gamma_{\axino\rightarrow \gamma \nu_i} = \frac{C_{aYY}^2 \alpha_{\rm em}^2}{128\pi^3} \frac{m_\axino^3}{\fa^2}  |U_{\tilde{\gamma}\tilde{Z}}|^2\sum_i \bfrac{\sqrt2 \VEV{\tilde{\nu}_i}}{v}^2 ,
 }
 where the photino-Zino mixing is given by
 \dis{
 U_{\tilde{\gamma}\tilde{Z}} = M_Z \sum_\alpha \frac{S_{\tilde{Z} \alpha} S^*_{\tilde{\gamma} \alpha} }{m_{\tilde{\chi}_\alpha}},
 }
 with the neutralino mixing matrix $S$. In the case of $M_1\ll M_2,\, \mu$, it is simplified to be
\dis{
\Gamma_\axino\simeq \frac{C_{aYY}^2 \alpha_{\rm em}^2}{128\pi^3} \frac{m_\axino^3}{\fa^2} \sum_i\xi_i^2.
}
Although axino can also decay to three neutrinos mediated by Z-boson, this mode is highly suppressed and negligible.

\begin{figure}[t]
  \begin{center}
  \begin{tabular}{c}
   \includegraphics[width=0.6\textwidth]{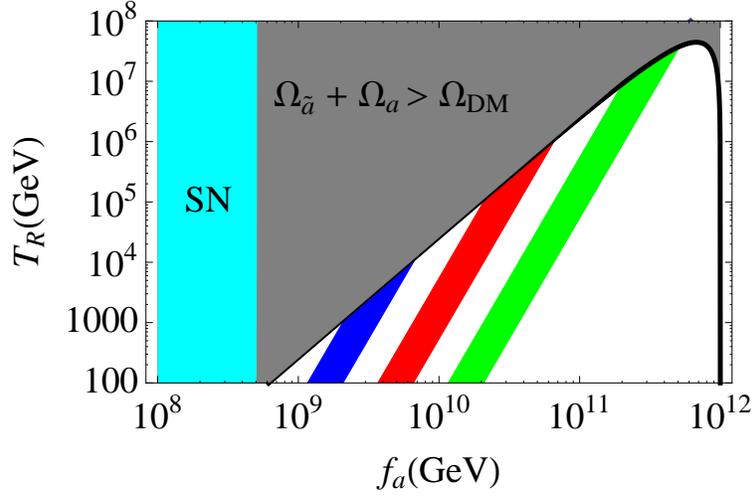}
   \end{tabular}
  \end{center}
  \caption{ The reheating temperature  $\treh$  versus $\fa$ for  given $\xi_i=10^{-5},10^{-4},10^{-3}$ (Blue, Red, Green) respectively  to explain X-ray line emission. The small value of $\fa<5\times 10^8\gev$ (cyan) is disallowed by the SN1987A. On the curved black line the thermally produced axino  and non-thermally produced axion (misalignment) can give correct relic density for dark matter. The upper region of the black line is disallowed due to the overabundance of axino and axion dark matter. }
\label{fig:TRfa}
\end{figure}

The X-ray emission line observed by the XMM-Newton can be explained with an appropriate lifetime and the relic abundance of axinos, if those satisfy the relation
\begin{equation}
\tau_\axino  =  \tau_{\rm DM} \bfrac{\abunda}{0.1},
\end{equation}
where $\tau_{\rm DM}$ is given in \eq{xray}. We note here that, even though the axinos are not the dominant component of dark matter, the enhanced decay rate can compensate to adjust the observed flux of X-ray line.

In Fig.~\ref{fig:TRfa}, we show the parameter space of  $\treh$  versus $\fa$  to explain the X-ray line emission for given R-parity violating parameters $\xi_i=10^{-5},10^{-4},10^{-3}$ (Blue, Red, Green) respectively. 
Here the small value of $\fa<5\times 10^8\gev$ (cyan) is disallowed by the SN1987A and the upper gray region is ruled out by the overabundance of the thermally produced axino and axions produced by the misalignment mechanism with an order unity misalignment angle~\cite{BaeHuhKim09}.
 
In the white region and on the black strip,
 the axino decay can explain the X-ray line emission with proper values of $\xi_i \gtrsim 10^{-5}$. 
On the black strip, axino constitutes whole dark matter for $ f_a < 10^{12}$ GeV, and axion contribution to dark matter is significant for $ f_a \simeq 10^{12}$ GeV. 
For example, for $\xi_i = 10^{-4}$ in the region along the red stripe,
the $3.5$ keV X-ray line emission can be well explained and, on the crossing point with the black strip, axinos explain total dark matter in the Universe as well as the X-ray line emission.
In the white region, the axino does not constitute all the component of dark matter the enhanced decay rate can properly adjust the required flux of X-ray. The rest of dark matter component must be compensated by another component of dark matter such as axions produced by another mechanism, e.g., the decay of heavier particles.

\begin{figure}[t]
  \begin{center}
  \begin{tabular}{c}
   \includegraphics[width=0.6\textwidth]{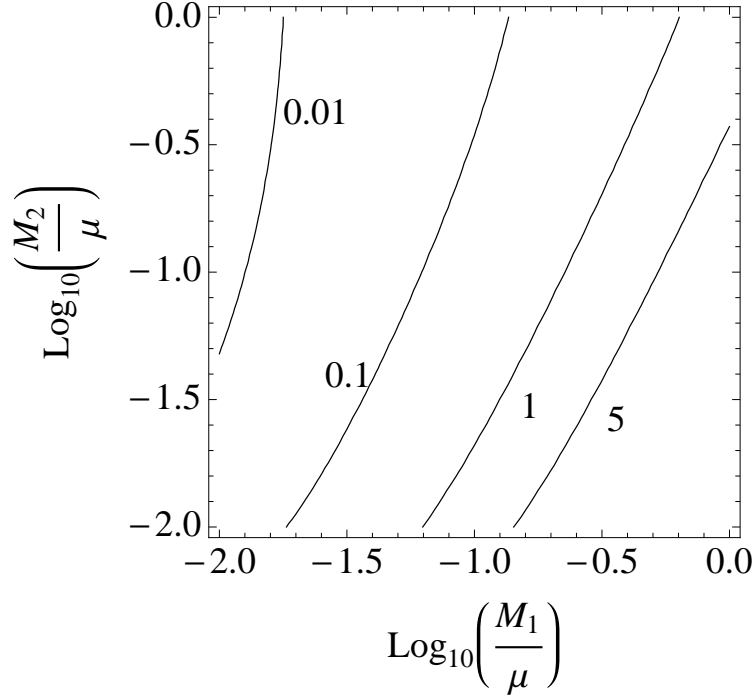}
   \end{tabular}
  \end{center}
  \caption{ The contour plot of $m_\nu / ( \mu \sum_i\xi_i^2)$ in \eq{mnu} in the plane of $(M_1/\mu, M_2/\mu)$. Here we used $\sin\beta\cos\beta=1/2$ for simplicity.}
\label{fig:xi}
\end{figure}

The relatively large $\xi_i$ can be inconsistent with the mass of neutrino from \eq{mnu}. 
In Fig.~\ref{fig:xi}, we show the contour plot of $m_\nu / ( \mu \sum_i\xi_i^2) $ in the plane of $(M_1/\mu, M_2/\mu)$. To explain the neutrino mass $m_\nu \lesssim 1 \ev$ with $\mu\sim 1\tev$
and $\xi_i \simeq 10^{-5}$, 
the value in the contour need to be smaller than $0.01$. That can be obtained when $M_1 \lesssim 0.01 \mu$ with slight dependence on $M_2$.
An interesting point to note is that the value of $\xi_i \gtrsim 10^{-5}$  implies light Bino $M_1 \lesssim 10\gev$ for $\mu=1 \tev$ from the upper bound for neutrino mass.

\section{Conclusion}
\label{conclusion}

Axino is a good candidate for dark matter. When its mass is around $7\kev$  and R-parity is broken bilinearly, the axino decays into photon and neutrino.
We studied this decaying axino warm dark matter in the light of the recent observation of X-ray line emission from the center of galaxy clusters and Andromeda galaxy observed by XMM-Newton. We find that the decaying axino can naturally explain the X-ray signal with/without additional component of dark matter. We note that the neutrino mass bound implies that the Bino mass is less than about $10$ GeV.

\section*{Note added}
As this work was being submitted, a paper~\cite{Kong:2014gea} appeared that also discusses
decaying axino dark matter as a source of the X-ray line signal.

\section*{Acknowledgments}
K.-Y.C. would like to  acknowledge the hospitality of Hokkaido University during his visit, where a part of this work was carried.
K.-Y.C. was supported by the Basic Science Research Program through the National Research Foundation of Korea (NRF) funded by the Ministry of Education, Science and Technology Grant No. 2011-0011083.



\def\prp#1#2#3{Phys.\ Rep.\ {\bf #1} #2 (#3)}
\def\rmp#1#2#3{Rev. Mod. Phys.\ {\bf #1}  #2 (#3)}
\def\anrnp#1#2#3{Annu. Rev. Nucl. Part. Sci.\ {\bf #1} #2 (#3)}
\def\npb#1#2#3{Nucl.\ Phys.\ {\bf B#1}  #2 (#3)}
\def\plb#1#2#3{Phys.\ Lett.\ {\bf B#1}  #2 (#3)}
\def\prd#1#2#3{Phys.\ Rev.\ {\bf D#1}, #2  (#3)}
\def\prl#1#2#3{Phys.\ Rev.\ Lett.\ {\bf #1}  #2 (#3)}
\def\jhep#1#2#3{JHEP\ {\bf #1}  #2 (#3)}
\def\jcap#1#2#3{JCAP\ {\bf #1}  #2 (#3)}
\def\zp#1#2#3{Z.\ Phys.\ {\bf #1}  #2 (#3)}
\def\epjc#1#2#3{Euro. Phys. J.\ {\bf #1}  #2 (#3)}
\def\ijmp#1#2#3{Int.\ J.\ Mod.\ Phys.\ {\bf #1}  #2 (#3)}
\def\mpl#1#2#3{Mod.\ Phys.\ Lett.\ {\bf #1}  #2 (#3)}
\def\apj#1#2#3{Astrophys.\ J.\ {\bf #1}  #2 (#3)}
\def\nat#1#2#3{Nature\ {\bf #1}  #2 (#3)}
\def\sjnp#1#2#3{Sov.\ J.\ Nucl.\ Phys.\ {\bf #1}  #2 (#3)}
\def\apj#1#2#3{Astrophys.\ J.\ {\bf #1}  #2 (#3)}
\def\ijmp#1#2#3{Int.\ J.\ Mod.\ Phys.\ {\bf #1}  #2 (#3)}
\def\apph#1#2#3{Astropart.\ Phys.\ {\bf B#1}, #2 (#3) }
\def\mnras#1#2#3{Mon.\ Not.\ R.\ Astron.\ Soc.\ {\bf #1}  #2 (#3)}
\def\nat#1#2#3{Nature (London)\ {\bf #1}  #2 (#3)}
\def\apjs#1#2#3{Astrophys.\ J.\ Supp.\ {\bf #1}  #2 (#3)}
\def\aipcp#1#2#3{AIP Conf.\ Proc.\ {\bf #1}  #2 (#3)}


\end{document}